\documentclass[11pt,twoside]{article}

\usepackage{asp2004}
\usepackage{epsf}
\usepackage{psfig}
\usepackage{lscape}

\begin{document}
\title{Accretion Disks, Jets and Blazar Variability}   
\author{Paul J.\ Wiita}   
\affil{Department of Physics and Astronomy, Georgia State University, Atlanta GA 30302-4106, USA}    

\begin{abstract} Although blazar variability is probably dominated by emission from
relativistic jets, accretion disks should be present in all
blazars.  These disks produce emission over most of the electromagnetic
spectrum; various unstable processes operate in those disks which will
lead to variable emission.  Here I summarize some of the most relevant disk mechanisms
for AGN variability. I also discuss some aspects of jet variability, focusing on
the possibility that ultrarelativistic jets of modest opening angle
can reconcile TeV blazar emission  with the many subluminal 
VLBI knots seen in those sources.  Finally, I present recently illuminated 
characteristics of optical microvariability of different classes of AGN 
which have important implications for the dominant processes involved. 
\end{abstract}


\section{Introduction}

Regardless of the exact fashion in which the jets which dominate blazar
spectra are launched (e.g.\ Vlahakis, these proceedings; Meier, these proceedings;
Hardee, these proceedings), 
all plausible models require them to either be anchored in
accretion disks or to emerge from the immediate vicinity of black holes which
are accreting.  Most modes of accretion produce significant fluxes from the
disks themselves, and most of those emissions will be variable.  There are two
 key questions 
with respect to accretion disks and blazar variability.  Can these disk emissions ever compete with
the jet emissions so as to make detectable contributions?  Can variations from
these disks occur on timescales short enough to be detected over the timespans
we are able to observe blazars?
In this brief review I concentrate on these questions but also
address a couple of specific points related to blazar jets. 
Czerny's (2004) recent review also covers many aspects of accretion disks that will be 
stressed here;
Czerny also provides nice discussions of other aspects of AGN variability 
not covered here and additional references.  
Key observations and basic theoretical points about disks are summarized in \S\S 2--4.
Aspects of blazar jets are discussed in \S 5 and results and implications of 
new optical microvariability 
studies of different AGN classes are noted in \S 6.

\section{Contributions from Accretion Disks to AGN Spectra}

The most direct evidence for accretion disk (hereafter AD) emission comes from 
the quasi-thermal big blue bumps sometimes seen in AGN spectra, which are
otherwise roughly characterized by power-law spectral energy distributions
in the IR--UV.  While these bumps are not
uncommon in quasars, their presence in blazars is
rare, presumably because the disk emission is usually swamped by the boosted
nearly power-law continuum.  New quasi-simultaneous multi-band observations of 
the blazar AO 0235+164 (Raiteri et al.\ 2005)
do provide such strong evidence for a big blue bump.  Additional recent evidence for AD
emission in several quasars comes from the detection of an optically
thick Balmer edge revealed in the polarized emission (e.g.\ Ton 202; Kishimoto et al.\ 2004).
Other types of direct observational evidence for flattened geometries,
presumably related to ADs, include the broad Fe K$\alpha$ lines seen in
Narrow Line Seyfert 1 galaxies (e.g.\ Reynolds \& Nowak 2003) and the variable
double peaked emission lines seen in a substantial fraction of
quasar spectra (e.g.\ Strateva et al.\ 2003).

The now fairly well accepted picture for accretion flows indicates that the relative
thickness of an AD and its extent toward its central black hole (BH)
are predominantly determined by its accretion rate, $\dot M$.  For accretion flows
where the accretion luminosity, $L$, is comparable to the Eddington limit,
 $L/L_E \sim 1$ ($L_E \simeq 1.3 \times 10^{46}  M_8 \ {\rm ergs \  s}^{-1}$,
with $M_{BH} = 10^8 M_{\odot} M_8$), the disk becomes geometrically and
optically thick (e.g.\ Paczy{\'n}ski \& Wiita 1980) and the inner edge of the
disk is inside the marginally stable orbit.  These bloated ``Polish donuts''
are  quite inefficient
radiators, since much of the ample quasi-thermal radiation they produce is trapped within the AD
and swept through the event horizon.
For flows where $L/L_E \sim 0.1$ the ``standard'' geometrically thin, 
but optically thick, AD picture (e.g.\ Shakura
\& Sunyaev 1973) basically holds.  For these, the most extensively studied
class of AD, the inner edge is at the marginally stable orbit
($r_{ms} = 6 GM_{BH}/c^2 \equiv 3 r_s$ for a Schwarzschild BH) and the efficiency
of mass to radiation conversion is high ($> 0.057$)
since most of the radiation produced in the AD escapes.  

At sufficiently
low accretion rates, producing $L/L_E < 0.01$, a two temperature flow 
is likely to develop (e.g.\ Rees et al.\
1982); here the ions in the infalling plasma can reach virial temperatures,
but their low densities mean that they fall into the BH before they can share their energy with the electrons, 
which do the bulk of the radiating.  There are various possible detailed structures
for such low ${\dot M}$ flows (e.g.\ Chakrabarti 1996),
but they are often generically called ADAFs 
(for Advection Dominated Accretion Flows; e.g.\ Chen et al.\ 1995). 
These flows are characterized by a transition radius, $r_t > r_{ms}$, inside of which
the flow is geometrically thick but optically thin, and outside of which it is 
physically thin but optically thick.  

The common view is that AGN
involve the first two categories of accretion flows with the
thin disk category most common.  On the other hand,  starved BHs, such
as the one in our Galactic nucleus, are of the last type.  Less well
accepted, but certainly possible, is the hypothesis that intermediate
luminosity AGN, such as radio galaxies and normal Seyfert galaxies, may also
have thin disks which terminate at $r \sim 100 r_s$ (see Czerny 2004 and
references therein).   It should be stressed that any particular galactic
nucleus may (and probably does) exhibit different accretion modes at different
times in its history and therefore can evolve from inactive, to 
one type of AGN, to another type of AGN.

\section{Timescales for Accretion Disk Variability}

If one adopts cylindrical coordinates, $(r,z,\phi)$, assumes a
quasi-Keplerian flow,  and expresses distances in terms of the
Schwarzschild radius, $r_s$, so that\\ $R \equiv r/r_s$, one finds that the dynamical
timescale in an AD is nearly the same as for a Keplerian orbit, and so is
\begin{equation}
t_{dyn} = 2 \times 10^3 R^{3/2} M_8 \ {\rm s}.
\end{equation}
Thus one should not observe
variations faster than
a few minutes for relatively low mass AGN (with $M_8 < 0.1$) or faster than
several hours for very massive AGN (with $M_8 > 10$) if there are no bulk relativistic
motions.  Such fast
variations associated with ADs would demand perturbations that affect a 
significant portion of the inner part of a disk and 
form and decay on orbital timescales.

Perhaps a more realistic timescale for many physical variations 
would be associated with the time needed
for a sound wave to be transmitted across a radial region of extent $r$
for a disk of thickness $h$,
$t_{sound} \simeq t_{dyn}(r/h)$;
this is roughly an order of magnitude greater than $t_{dyn}$ for thin AD
models.

Another key timescale on which one would expect to see changes in 
 any AD is its thermal timescale, $t_{th}$,
defined as the ratio of the disk's internal energy to the heating or cooling rate.
If, for simplicity, we assume the Shakura--Sunyaev $\alpha$-disk parameterization,
where the shear stress is related to the total pressure by the expression 
$T_{r \phi} \equiv \alpha P$, then the thermal timescale is
\begin{equation}
t_{th} = t_{dyn}/\alpha = 2 \times 10^4 \alpha^{-1}_{-1} R^{3/2} M_8 \ {\rm s},
\end{equation}
where a typical value of $\alpha = 0.1 \alpha_{-1}$ has been used.  These
thermal timescales are of the order of days for $10^8 M_{\odot}$ BHs and are
comparable to $t_{sound}$.

One of  the most interesting longer timescales is 
related to the rate at which matter flows through the AD.  This viscous timescale
is simple to estimate if we stick to the $\alpha$-disk approximation (e.g.\ Czerny 2004):
\begin{equation}
t_{visc} \simeq t_{th}\Bigl(\frac{r}{h}\Bigr)^2 = 2 \times 10^6 \alpha^{-1}_{-1} 
\Bigl(\frac{r}{h_{-1}}\Bigr)^2 R^{3/2} M_8 \ {\rm s},
\end{equation}
where we have assumed a typical value of $h = 0.1 h_{-1}$. Thus, substantial
 variations in flux over
months to years (for $M_8 \sim 1$), could be related to viscous processes.

If there is an ADAF, with a
hot inner flow and a cold outer disk, or if there is a strong X-ray producing corona sandwiching a 
standard thin disk, then the transition radius, $r_t$, may slowly change with
time, either by evaporation or by a significant outflow of gas.  
The key question becomes: how fast can the cold disk be removed?
In either event, the disk temperature must rise to roughly the virial temperature,
and enough energy to perform this  heating must be stored up in the disk.
For an $\alpha$-disk, this gives $t_{evap} \equiv t_{visc}$, but more generally 
and realistically it
can be shown that (Czerny 2004)
\begin{equation}
t_{evap} \simeq 1000 \Bigl(\frac{r_t}{100 r_s}\Bigr)^2 {\dot m}_{-1} M_8 \ {\rm yr},
\end{equation}
where the accretion rate, ${\dot m}_{-1}$ has been normalized to 0.1 of that 
needed to produce $L_E$: ${\dot m} \equiv {\dot M}/{\dot M_E}$.  For
 reasonable values of $r_t$, such variations could
only be relevant (over the decadal periods we can observe AGN) for relatively
low values of BH mass or ${\dot M}$, such as those associated with Seyferts.  But given the
massive galactic hosts of most blazars (at least for $z > 0.5$; Kotilainen et al.\ 2005) 
and the nearly linear relation
between galactic halo mass and that of the central BH, 
this timescale is likely to exceed $10^3$ yr for blazars.

The longest interesting timescale, $t_{fuel}$, is that over which
the accretion rate changes through  differences in the amount of gas available to the
BH.  While an occasional gas bolus may be come from a star disrupted
in the vicinity of the BH, the fundamental availability time is comparable
to the lifetime of the AGN phenomenon itself.
Fueling should be driven by processes
related to galactic mergers or harassment and will typically be $t_{fuel} \sim 10^7$--$10^8$ yr.

\section{Physical Mechanisms for Accretion Disk Variability}

The radiation pressure instability has long been known to afflict $\alpha$-disks 
(Shakura \& Sunyaev 1976).  Since AGN ADs should be radiation pressure dominated over substantial
ranges in $r$ and ${\dot M}$ this instability should yield fluctuations
in emission, although it is unlikely to disrupt the disk.  Recent simulations of
these variations produce substantial flares on timescales of
 $t_{visc}(\sim 100 r_s)$ (e.g.\ Teresi et al.\ 2004).  This class of variation
may be the appropriate explanation for fluctuations in the microquasar GRS 1915+105 
(Teresi et al.\ 2004) over minute timescales.  If scaled to AGN masses, 
the radiation pressure instability
may yield big AD outbursts from the X-ray through IR over years to decades.

The magneto-rotational instability (MRI) is now commonly agreed to be present in all ADs
(e.g.\ Balbus \& Hawley 1991).  The MRI rapidly amplifies seed magnetic fields  
and is very likely to provide the dominant viscosity
in most ADs, providing an effective, but fluctuating, $\alpha \sim 0.01$--$0.1$ (e.g.
Armitage 1998; Miller \& Stone 2000).  These MRIs will produce strong turbulence which can yield significant
changes in both the production of
heat energy and effective $\dot M$ on rapid (a few $t_{dyn}(r_{ms})$) timescales, 
or hours for a $10^8 M_{\odot}$ BH, as shown
by recent simulations (Armitage \& Reynolds 2003).
While they should also produce some disk clumping and some larger variations on longer
timescales, the MRI is unlikely to destroy the disk.  Changes in line profiles associated with
ADs also may be produced by MRI. 

The double-humped emission lines detected in $\sim 10$\% of AGN
spectra (e.g.\ Strateva et al.\ 2003) are frequently observed to show
changes in the relative heights of the red and blue humps.  These
changes, and the shapes of the features themselves, are best explained by spiral shocks in ADs
(Chakrabarti \& Wiita 1994).  
The perturbation of
a disk by a smaller BH in the vicinity of the AD
can efficiently drive these shocks.  These spiral shocks can also produce factor of
a few variations in the AD emission on the orbital timescales of the perturbers, 
which, however, would
typically be decades
or longer (Chakrabarti \& Wiita 1993). 

Because ADs are almost certainly magnetized, there is a high probability that they are sheathed
by coronae, which are certain to contribute substantially to the X-ray emission and should 
have some role to
play in other bands.  While understanding of coronal structure formation above
ADs is still a long way off, simulations of MRI do illustrate the formation of large loops
over the course of several $t_{dyn}$ (e.g.\ Miller \& Stone 2000).  These coronal structures
probably play a significant part in the X-ray emission and variability of Seyfert galaxies,
but their importance for quasars and blazars is much less clear.  The variations in energy that can 
be released from coronal flares are probably too small to be detected, particularly outside
the X-ray band, in that
individual flares are  limited in their powers
(e.g.\ Krishan et al.\ 2003).  Still, large groups of coordinated flares, produced by an 
avalanche mechanism or other form of
self-organized-criticality (e.g.\ Mineshige et al.\ 1994) can produce the optical power-spectrum 
density observed in many AGN (Xiong et al.\ 2000).

\section{Jets and Blazar Variability}

As this key topic is the focus of many other papers in this volume, 
including those by M.\ Aller, B{\"o}ttcher, Georganopoulos, Joshi, and Kazanas,
here I make only a few
general points. My focus is on two questions: Is there a need for coherent emission
from blazar jets?  Can subluminal
VLBI motions be reconciled with TeV emission from blazars?  

Clearly, relativistic shocks
propagating down jets can explain much of the gross radio through optical variations via
boosted synchrotron emission.  Turbulence, instabilities, and magnetic inhomogeneities
can probably explain the bulk of rapid variations.  The inverse Compton (IC) mechanism, invoked
either through synchrotron self-Compton, external Compton, or decelerating jets can explain
particular observed high energy variations with respect to low energy ones. Still, no relatively
simple model seems able to explain most broad band observations, and perhaps multiple
sources of IC seed photons are needed.
 
Compact radio sources with intrinsic brightness temperatures, $T_{B,int} > 10^{11}$K 
exceed the self-absorbed source IC catastrophe limit (Singal \& Gopal-Krishna 1985).
Early observations of intraday variability (IDV) at cm-wavelengths seemed to imply $T_B \sim 10^{21}$K. 
 Bulk relativistic motions with {\it very} high Lorentz factors ($\Gamma\equiv (1-\beta^2)^{-1/2} \sim 10^3$)
are required to bring these into accord, as a factor of $\Gamma^3$ converts the intrinsic $T_B$ into
the observed one.  While these huge $\Gamma$ factors prevent the production of too many X-rays, they do so
at the cost of very low synchrotron radiative efficiencies, and thus demand very high jet powers.
This problem seemed to go away when later observations showed that
most IDV is due to refractive interstellar scintillation (e.g.\ Kedziora-Chudczer et al.\ 2001);
then $T_{B,int} \sim 10^{13}$K so modest $\Gamma$'s ($\sim 30$) remove the catastrophe.  
But a recent unpublished claim
that the blazar J1819+3845 shows diffractive scintillation with a size $< 10 \mu$as (Macquart \& de Bruyn
2005) implies
$T_{B,int} \gg 10^{14}$K; if true, $\Gamma > 10^3$ is needed to allow incoherent synchrotron emission.

Coherent radiation induced by strong Langmuir turbulence in AGN jets can produce the needed huge
$T_B$'s without requiring such extreme Lorentz factors (e.g.\ Krishan \& Wiita 1990), but such
coherent models (implicitly)
assume $\nu_{plasma} > \nu_{cyclotron}$.  Recently Begelman et al.\ (2005) have argued that
the opposite is much more likely to be the case in blazar jets.  Begelman et al.\ (2005)
propose that many transient small-scale magnetic mirrors can be produced from hydromagnetic
instabilities, shocks or turbulence. The electrons accelerated along these converging flux tubes can
quite naturally produce the population inversion needed for a cyclotron maser.  Fundamentally, this
class of maser is pumped by turning kinetic and magnetic energy into ${\vec j}\cdot{\vec E}$ work.
If future observations do wind up supporting the need for coherent emission processes in blazars,
this cyclotron maser mechanism appears to be quite promising.

\begin{figure}
\plotone{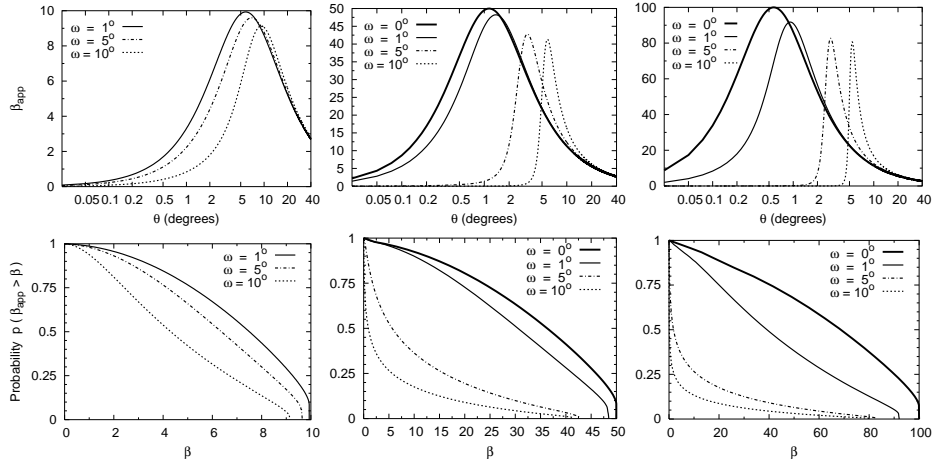}
\caption{The upper panels give distributions of $\beta_{app}$ vs.\ $\theta$ for $\Gamma = 10, 50$ and $100$.
Results for jet opening angles, $\omega = 0^{\circ}, 1^{\circ}, 5^{\circ}$ and $10^{\circ}$ are labeled.
The lower panels show the cumulative probabilities for $\beta_{app} > \beta$ for the same values of
$\Gamma, \theta$ and $\omega$.  From Gopal-Krishna et al.\ (2004); copyright, AAS}
\end{figure}

The only semi-direct probe of extragalactic jet speed comes from VLBI knot apparent motions,
and quite surprisingly, $> 30\%$ of these turn out to be subluminal ($\beta_{app} < 1$)
for TeV emitting blazars (Piner \& Edwards 2004).  On the other hand, to avoid excessive
photon-photon collisions, these TeV blazars 
require ultrarelativistic jets (Krawczynski et al.\ 2002) with $15 < \delta < 100$;  $\delta =
[\Gamma(1-\beta {\rm cos}\theta)]^{-1}$, with $\theta$ our viewing angle to the jet.  
Taking into account IR background absorption strongly implies $45 < \delta$ (Kazanas, these proceeding).  And, while rare
(Lister, these proceedings) some $\beta_{app} > 25$ components are seen in EGRET blazars (Piner et al.,
these proceedings).

Several ways have been proposed to reconcile the high $\delta$ factors with low apparent
speeds.  The leading idea is that jets might have fast cores (spines) which give rise to the $\gamma$-ray 
emission via IC while
slower outer layers (sheaths) produce the radio knot emission (e.g.\ Ghisellini et al.\ 2005).  
Alternatively, jets could
rapidly decelerate between sub-pc ($\gamma$-ray) and pc (VLBI knot) scales (Georganopoulos, Kazanas,
these proceedings).
Also, a few such cases could arise from viewing angles to within $\sim 1^{\circ}$, but there are too many
slow knots for this to be the main explanation.  Finally, it must be recalled that some of the 
motions may reflect pattern, not physical, speeds (Hardee, these proceedings).  

We (Gopal-Krishna et al.\ 2004) have recently shown another way to reconcile the very high
$\Gamma$s preferred by IC and variability arguments with the low $\beta_{app}$ values seen
in TeV blazar jets.  Instead of assuming that
the jets are cylindrical as was done in all previous models, we allow them to have small opening 
angles, $\omega$.
The resulting slightly different viewing angles to the finite jet 
emitting region require the computation of
weighted fluxes and weighted $\beta_{app}$ values.  
The apparent transverse speeds peak at lower values and at significantly higher
$\theta$ than for cylindrical jets (upper panels, Fig.\ 1).  We find many
sources with $\beta_{app} < 1$ and only a few with $\beta_{app} \sim \Gamma$ if  $\Gamma > 25$ and
$\omega > 5^{\circ}$ (lower panels, Fig.\ 1).  

\section{No Fundamental Difference Between Blazars and All Other AGN?}

We have recently completed an extensive monitoring campaign of optical microvariability 
for a group of 25 powerful AGN (Gopal-Krishna et al.\ 2003; Sagar et al.\ 2004;
Stalin et al.\ 2004a,b).  This program involved sets, matched
in optical luminosity and redshift, of: radio-quiet quasars (RQQs); lobe-dominated radio-loud
quasars (LDQs); core-dominated radio-loud quasars (CDQs); and BL Lacertae objects (BLLs).
In the standard unified scheme, the first two groups should not be affected by beaming, but the
latter two would be called blazars.  Although there had been various earlier microvariability
studies also using CCD differential photometry of most of these different classes, 
discrepant results were reported, particularly for RQQs.  I believe all earlier results were suspect,  
because of poor choices of objects, comparison stars,  observing techniques, or data reductions.
We are confident that our observing techniques, which always involved at least 4 hours of
frequent monitoring per source per night, along with our careful reductions, have often 
provided highly significant detections of variations as small as $\sim 0.01$ mag 
over the course of several hours.

One main result of this work is to confirm that BL Lacs (and the one high-polarization
CDQ in our sample) show a high duty cycle for microvariability ($\sim$ 60\%).  These blazars
showed variations of up to $\sim 0.14$ mag in the course of a single night (Sagar et al.\ 2004).  

A more
important result was the first clear detection of microvariability on several nights for different RQQs
(Gopal-Krishna et al.\ 2003; Stalin et al.\ 2004a).
Furthermore, the duty cycles for RQQs, LDQs and (low-polarization) CDQs were all 
$\sim$ 20\% (Stalin et al.\ 2004a,b).  All of the observed microvariations for these non-blazars 
ranged between 0.01 and 0.03 mag. 
The properties of the microvariability in RQQs and LDQs were indistinguishable 
in our studies, which would be unexpected if these optical variations were arising in
jets, in that the RQQs were all very weak in the radio band.  Our original motivation
for searching for RQQ microvariability was the expectation that, if any were found,
it could be logically attributed to AD fluctuations.

Still, all of these new results can be understood in terms of the standard beaming picture, where
the BLLs and high-polarization CDQs are viewed at small angles to the line-of-sight and the
RQQs and LDQs are viewed at larger angles.  The Doppler enhancement of the amplitude of the variations
and the Lorentz time-dilation for sources viewed at small $\theta$
together can provide for the higher duty cycles and more
powerful amplitude distributions of microvariability for the blazars, even if their rest-frame
characteristics are identical to those of the non-blazars (Gopal-Krishna et al.\ 2003; Stalin
et al.\ 2004a).  

One possible explanation of this microvariation unification is that all these classes
of AGN do possess jets, which, on sufficiently small scales, yield fluctuations which are seen in
the optical; however, in RQQs these jets are somehow quenched before significant radio emission is 
produced.  Another possibility is that these fluctuations do originate in the ADs, and are
seen unbeamed from non-blazars; in the case of blazars these fluctuations propagate into jets 
launched from the ADs and are amplified by the relativistic motion.  Additional observations will
be needed to distinguish between these hypotheses.

\section{Conclusions}

Accretion disks exist in all AGN, including blazars.  They can contribute noticeable
amounts to the fluxes emerging from blazars, at least in the IR--X-ray bands.  
Since ADs must vary on timescales we
can observe, some fraction of detected blazar variability can arise from ADs.
There are 
several ways to reconcile low VLBI knot speeds with high Lorentz factors in TeV blazars;
if several degree opening angles are present, no variations in jet speed are necessary to
do so.
Microvariability of RQQs and blazars can be interpreted
as arising from the same process; this 
either implies
that jets are present in all AGN or that variations produced in a disk may
be carried into, and amplified by, jets.
 

\acknowledgements I thank my collaborators,
 G.\ Bao, P.\ Barai, S.\ K.\ Chakrabarti, S. Dhurde, Gopal-Krishna,  V.\ Krishan, S.\ Ramadurai,
R.\ Sagar, C.\ S.\ Stalin
and Y.\ Xiong.
I am most grateful for hospitality at 
Princeton University and acknowledge support from NASA, 
NSF and RPE funds at GSU.

%

\end{document}